\documentclass{article}
\usepackage[T1]{fontenc}
\usepackage[utf8]{inputenc}
\usepackage[]{ismir} 
\usepackage{ismir} 
\usepackage{amsmath,cite,url, amssymb}
\usepackage{graphicx}
\usepackage{color}
\usepackage{booktabs}
\usepackage{MnSymbol}
\usepackage{multirow}

\newcommand\extrafootertext[1]{%
    \bgroup
    \renewcommand\thefootnote{\fnsymbol{footnote}}%
    \renewcommand\thempfootnote{\fnsymbol{mpfootnote}}%
    \footnotetext[0]{#1}%
    \egroup
}

\title{L\lowercase{i}LAC: A Lightweight Latent ControlNet for Musical Audio Generation}

\twoauthors
  {Tom Baker$^{1,2^*}$} {$^1$University Of Manchester \\ \tt\small{tom.baker@manchester.ac.uk}}
  {Javier Nistal$^2$} {$^2$Sony CSL - Paris \\ \tt\small{javier.nistal@sony.com}}

\def\authorname{T. Baker and J. Nistal}

\usepackage[bookmarks=false,pdfauthor={\authorname},pdfsubject={\pdfsubject},hidelinks]{hyperref}

\begin{document}

\maketitle
\begin{abstract}
Text-to-audio diffusion models produce high-quality and diverse music but many, if not most, of the SOTA models lack the fine-grained, time-varying controls essential for music production. ControlNet enables attaching external controls to a pre-trained generative model by cloning and fine-tuning its encoder on new conditionings. However, this approach incurs a large memory footprint and restricts users to a fixed set of controls. We propose a lightweight, modular architecture that considerably reduces parameter count while matching ControlNet in audio quality and condition adherence. Our method offers greater flexibility and significantly lower memory usage, enabling more efficient training and deployment of independent controls. We conduct extensive objective and subjective evaluations and provide numerous audio examples on the accompanying website.\footnote{\url{https://lightlatentcontrol.github.io}}

\end{abstract}

\extrafootertext{$^*$Research undertaken while an intern at Sony CSL - Paris}
\section{Introduction}\label{sec:introduction}
With the rise of generative models, new challenges have emerged in the field of human-computer interaction, especially in how users interact with these systems \cite{XAI}. This issue is particularly significant in domains of artistic expression, such as music creation, where users need interfaces that allow for both high-level control over abstract concepts and precise manipulation of low-level details. Achieving this balance between creative freedom and technical control is critical for musicians and composers when working with generative systems.

Generative models for music have explored various control mechanisms to bridge the gap between user intention and machine output \cite{musiclm,audioldm,stableaudio, musicgen, diffariff}. However, there is no clear consensus on which control modalities or signals are most effective. Currently, one of the most common methods for interacting with generative music systems is through text input\cite{musiclm,audioldm,stableaudio}, which leverages shared embedding spaces for text and audio information\cite{clap}. While this approach has facilitated significant advancements, it lacks the fine-grained control required for detailed music production.

In response to these limitations, some studies have experimented with conditioning generative models on time-varying signals, such as pitch or dynamics\cite{musicgen,jukedrummer}. These provide more specificity but are still constrained by a key issue: control signals are typically required during the training process of the generative model. Once the model has learned to respond to these inputs, the control mechanisms become fixed and inflexible.

Modular approaches like ControlNet \cite{controlnet} enable flexible, post-hoc control of image generative models and have recently been adapted to music generation \cite{musiccontrolnet,ditmusic}. While effective, these techniques come with important limitations: memory-intensive clones of the network for each new control \cite{controlnet}, or rigid multi-control schemes \cite{unicontrolnet}.

In this work, we adapt ControlNet’s framework to musical audio generation, introducing a modular architecture that replaces cloned encoder blocks with lightweight convolutional layers. While similar parameter-efficient approaches exist in computer vision\cite{controlnet-xs}, our method is the first to demonstrate this for music, enabling flexible training of multiple independent control models (e.g., chroma, chords) without retraining the backbone. By decoupling controls into task-specific modules, users can deploy only the necessary conditions during inference, often with lower memory overhead than a single traditional ControlNet branch. Crucially, our evaluations—spanning objective metrics (FAD, APA) and subjective listening tests—show that this streamlined design achieves performance comparable to ControlNet in audio quality and condition adherence, establishing a practical balance between flexibility and fidelity for music generation workflows.

\section{Related Work}\label{sec:related_work}

Controls for music generation models encompass diverse input modalities. Text prompting is widely used \cite{musiclm,mousai,audioldm,noise2music,jen1,audioldm2,stableaudio}, and recent joint text-audio embeddings allow zero-shot control without paired data \cite{mulan,clap,laion}, enabling edits like “make this piece of music more happy” \cite{musicmagus,zeroshot}. However, text can be ambiguous and less suited for precise control in music production.
Finer-grained controls, such as melody \cite{musicgen}, rhythm and dynamics \cite{jukedrummer}, or timbral features \cite{drumgan, drumganvst}, offer more precision. Multimodal inputs like images or video also expand creative possibilities \cite{makeanaudio}.
Audio-based conditioning is particularly effective for tasks like accompaniment generation and style transfer \cite{drumcontrol,bassnet,stemgen,bassdiffusion,multisource,composablereps,simulsep}, with models like Diff-A-Riff \cite{diffariff,dar2} and SingSong \cite{singsong} leveraging input audio to guide generation.
Control integration strategies vary—from training-time conditioning to inference-time guidance \cite{inferencetimeguidance} or latent optimization \cite{ditto,ditto2}. Inspired by ControlNet, recent methods introduce auxiliary networks for control \cite{musiccontrolnet,ditmusic}, while others, like Sketch2Sound \cite{sketch2sound}, explore lightweight alternatives.
Despite this progress, optimal strategies remain unclear. ControlNet-style designs offer modularity but are often resource-intensive or inflexible. To address this, we propose a new lightweight and modular architecture that retains ControlNet’s strengths with improved efficiency.

\section{Background}\label{sec:background}
This work builds on ControlNet \cite{controlnet} (see Sec.~\ref{sec:cnet}), a framework for introducing post-hoc controllability into pre-trained generative models. While our architecture is generalisable to any generative model, we utilise Diff-a-Riff \cite{diffariff} (see Sec.~\ref{sec:dar}) as the \emph{backbone} model throughout this paper. In the following sections, we provide an overview of these two architectures, laying the groundwork for the proposed methodology.

\subsection{Diff-A-Riff}\label{sec:dar}
Diff-a-Riff\cite{diffariff} is a Latent Diffusion Model (LDM) designed to generate high-quality individual musical stems that align with a user-provided musical audio sample, denoted \emph{Context}. The model employs a Consistency Autoencoder (CAE)\cite{music2latent} to compress raw audio into compact latent representations and utilises an Elucidated Diffusion Model (EDM)\cite{edm} framework. The CAEs latent audio codec reduces 48~kHz audio to a 64-dimensional encoding at $\sim$12~Hz. Generation can be controlled via audio references, text prompts, or interpolations of both, facilitated by a shared CLAP embedding space \cite{clap,laion}. For further details, refer to the original paper\cite{diffariff}.

\subsection{ControlNet}\label{sec:cnet}
ControlNet \cite{controlnet} introduces a method to augment large pre-trained text-to-image diffusion models with new controls. It achieves this by freezing the parameters of the original model, or \emph{backbone}, and introducing a so-called \emph{adaptor branch}—a trainable copy of the backbone's encoding layers. This branch processes both original inputs and new conditional signals, feeding activations back through zero-initialised convolutions while reusing only the original training objective.

The decoupled architecture allows for conditioning with limited specialised data, enabling diverse controls (edges, depth maps, segmentation, poses) without compromising the pretrained backbone's capabilities.

ControlNet has been successfully adapted to musical audio, providing time-frequency controls like pitch or loudness \cite{musiccontrolnet, ditmusic}. Below, we provide a brief overview of the details relevant to this paper.

\subsubsection{Architecture}\label{sec:cnetarch} 
The architecture is displayed in Fig.~\ref{fig:cnetlilac}. For layer $l$, we utilise both a frozen encoder block $\mathcal{F}_l(x_{l-1}, e)$ and its cloned adaptor block counterpart $\mathcal{G}_l(\hat{x}_{l-1},e)$.  Here, $x_{l-1}$ and $\hat{x}_{l-1}$, represent the outputs from the previous frozen and control layers, respectively, while $e$ denotes the backbone's original conditional embeddings. Using zero convolutions $\mathcal{Z}_s$, the skip connection $s_l$ is computed as:

\begin{equation}\label{eqn:cnet_skip}
    s_l = \mathcal{F}_l(x_{l-1},e) + \mathcal{Z}_s(\mathcal{G}_l(\hat{x}_{l-1},e))
\end{equation}

The input to the adaptor branch $\hat{x}_0$ is derived from the noised input tensor $x_0$ and the new conditional $c$ through the input zero convolution: $\hat{x}_0 = x_0 + \mathcal{Z}_{in}(c)$.

ControlNet \cite{controlnet} demonstrates that using cloned encoders from the backbone model is critical for effective control signal integration, as randomly initialised convolutions compromise condition adherence, particularly when text conditioning is misaligned. Additionally, zero convolutions—used to introduce the control signal via skip connections—gradually introduce the signal during training, improving stability and output quality.

\section{LiLAC}\label{sec:method}
In this section, we introduce LiLAC and detail how its architecture, conditioning mechanisms, and training methodologies diverge from ControlNet.

\subsection{Proposed Architecture}\label{sec:layers}

\begin{figure}
    \centering
    \includegraphics[width=0.97\linewidth]{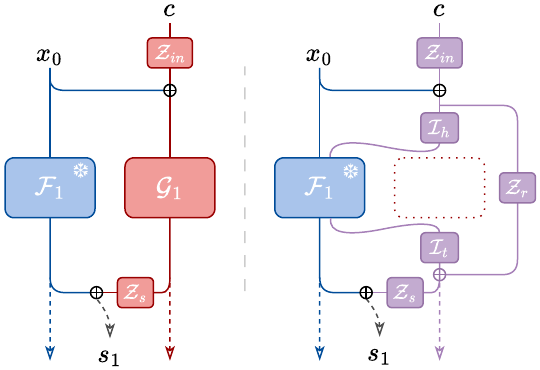}
    \caption{ControlNet~\cite{controlnet} (left) and LiLAC (right) adaptor blocks ($l=1$). The noisy input signal is denoted as $x_0$, and conditional signal as $c$. The frozen encoder block is denoted $\mathcal{F}_1$, and its cloned copy $\mathcal{G}_1$. Identity $\mathcal{I}$ and zero $\mathcal{Z}$ convolutions, along with the skip connection pathway $s_1$ are also illustrated.}
    \label{fig:cnetlilac}
\end{figure}

In Fig.~\ref{fig:cnetlilac}, we depict the basic block of LiLAC's architecture. Instead of cloning the backbone's encoder, LiLAC performs a second pass through each of the frozen encoder blocks, wrapping these by smaller convolutional layers. Specifically, we introduce three layers per block: a \emph{head} layer before the frozen block, a \emph{tail} layer after the frozen block, and a \emph{residual} connection to preserve condition information as it passes through the frozen block.

Formally, we replace the cloned encoder block $\mathcal{G}_l(\hat{x},e)$ in (\ref{eqn:cnet_skip}) with
\begin{equation*}
    \mathcal{G}_l(\hat{x},e) \approx \underbrace{\mathcal{I}_t}_\emph{tail}(\mathcal{F}_
l(\underbrace{\mathcal{I}_h(\hat{x})}_\emph{head},e)) + \underbrace{\mathcal{Z}_r(\hat{x})}_\emph{residual},
\end{equation*}
where $\mathcal{I}$ represents the identity convolutions used as the \emph{head} and \emph{tail} layers (see Sec.~\ref{sec:identity}), and $\mathcal{Z}_r$ denotes the zero convolution used as the \emph{residual} connection.  

While other multi-control methodologies have proposed reintroducing the condition into each block~\cite{musiccontrolnet}, we found empirically that this approach does not improve condition adherence and adds redundant parameters.

\subsubsection{Identity Convolutions}\label{sec:identity}

As discussed in Sec.~\ref{sec:cnetarch}, ensuring the adaptor branch leverages the backbone model's knowledge while gradually introducing the conditional signal during training is crucial. We achieve this by initialising the adaptor pathway to mirror the backbone encoder through identity convolutions $\mathcal{I}$, in the \emph{head} and \emph{tail} layers, and zero convolutions, $\mathcal{Z}$, in the \emph{residual} layer.

We initialise any $n$-dimensional Identity convolution kernel biases to 0 and the weights $W_\mathcal{I}[k_1,...k_n,,i,j]$ as:
\begin{equation*}
    W_\mathcal{I}[k_1,...k_n,i,j] = \begin{cases}
        1 & k_1,...k_n=\frac{K-1}{2}, \text{and}\ i=j\\
        0 & otherwise
    \end{cases}
\end{equation*}

Where $k_1,...k_n$ index spatial positions in the kernel, and $i,j$ index input and output channels. $K$ is the kernel size and must be odd; in our case, all convolution kernels are size 1 to maintain a lightweight network.

\subsection{Conditioning Signals}
As outlined in Section \ref{sec:layers}, the condition $c$ is introduced to the adaptor branch by adding it to the noised input $x_0$. However this requires dimensional alignment between both tensors. For $n$-dimensional input data with $C$ channels, the noised input is represented as $x_0\in \mathbb{R}^{C\times k_1\times...\times k_n}$. Conditions are provided as feature maps in the form $c \in \mathbb{R}^{N\times k_1\times...\times kn}$, where $N$ is the number of condition channels. To align the condition with the input tensor, the initial zero convolution $\mathcal{Z}{in}$ also maps the condition's channels from $N$ to $C$ before they are combined.

\section{Experimental Setup}\label{sec:experiments}

\subsection{Data}
In this work, we use the same dataset from Diff-A-Riff \cite{diffariff}, comprising 12,000 multi-track recordings, randomly split into 1 million 10-second segments. During training, the model receives one track as the target and a random combination of the remaining tracks as musical \emph{Context}.

\subsubsection{Conditions}
In addition to the original training data, we pre-extract the required conditioning control signals from each audio track as specified below. For evaluation purposes, we select two distinct signals to address different use cases: chroma pitch class conditioning and chord conditioning.

\vspace{2mm}
\noindent \textbf{Chroma.} A chromagram is extracted from the target single track using the librosa library \cite{librosa}. The 12 chroma bins are placed along the channel dimension, and the time scale is downsampled to match the frame rate of the latent audio codec used by the backbone model ($\sim$~11.7~Hz) \cite{music2latent}. 

\vspace{2mm}
\noindent \textbf{Chord.} We extract chord symbols for each frame from mixed multi-track audio using Deep12MIR \cite{deep12} and convert them into a 12-dimensional chroma-like format. Each chord is encoded as a vector where pitch classes in the chord are assigned 1, the root note 2, and remaining entries 0. This format supports all chord symbols, with harmonically related chords receiving similar embeddings, and facilitates the use of arbitrary chord shapes absent in the training data.

\subsection{Training details}
We follow the backbone's training methodology~\cite{diffariff}. Conditions \emph{Context}, $e$, and $c$ are all independently dropped out with a 50\% probability during the training of the adaptor branch \cite{controlnet}, encouraging the model to use the new condition, $c$ as a replacement for missing information and enabling Classifier Free Guidance (CFG) \cite{CFG} at inference. All models are trained for 2 days on a single NVIDIA RTX 3090 GPU with a batch size of 128. We use AdamW\cite{adamw} optimizer with a base learning rate of $10^{-4}$ and a cosine annealing learning rate scheduler with linear warmup\cite{cosanneal}.

\subsection{Objective Evaluation}
In the following section, we present two objective evaluation strategies: one comparing our methodology against baseline models, and the other analysing condition conflicts and their impact on condition adherence.

\subsubsection{Metrics}
We employ a subset of the metrics used in the backbone paper\cite{diffariff}: \emph{Fréchet Audio Distance} (FAD) \cite{fad} for audio quality and diversity; \emph{Audio Prompt Adherence} (APA) \cite{apa} and \emph{CLAP Score} (CS) for context and text adherence, respectively. For evaluating adherence to the LiLAC-provided chromagram conditioning we calculate the \emph{chroma Mean Squared Error} (cMSE) between input and output chromagrams. All metrics are computed by averaging across five sets of 500 samples. Distribution-based metrics such as FAD and APA are measured using the CLAP embedding space \cite{clap,laion} against a background dataset of 5,000 real audio samples.

\subsubsection{Baselines}
We compare LiLAC against four baseline configurations: the backbone model in three conditioning setups—totally unconditioned (\emph{Diff-a-Riff*}), with CLAP (\emph{Diff-a-Riff}), and with both CLAP and context (\emph{Diff-a-Riff + Context})—as well as the original \emph{ControlNet} architecture applied to the backbone. 

\subsubsection{Comparison and Ablation}\label{sec:cab}
With the aim of keeping the architecture as light as possible, we conduct an ablation study for each of the three LiLAC adaptor layers described in Section \ref{sec:layers}-\emph{Head} (H), \emph{Tail} (T), and \emph{Residual} (R). We evaluate their performance relative to the parameter overhead (see Tab.~\ref{tab:ablation}). Additionally, we test a model variant denoted \emph{LiLAC*} that avoids a second pass through the backbone model by directly inserting the new adaptor layers as part of the backbone's forward pass, similar in idea to previous work \cite{sketch2sound} (Although notably we don't further fine-tune our backbone model).

For the objective evaluations, we deliberately employ raw chromagram conditioning due to its high information density, enabling us to establish a more accurate upper bound on the model's conditional adherence performance. Since each audio signal produces only a single chromagram, this allows for precise and unambiguous objective comparisons using the \emph{cMSE}. In contrast, chord symbols can have multiple valid interpretations for a single stem, and as there is no unambiguous single way to evaluate adherence across a variety of instrument classes, we reserve these for the subjective test (see Sec.~\ref{sec:sa}).

\subsubsection{Conflictive Conditioning}\label{sec:cc}

Next, we evaluate how LiLAC and ControlNet-style models respond to conflicting CLAP and chroma conditionings (e.g., a solo violin CLAP embedding with a polyphonic chroma condition). Certain conditions, especially those extracted directly from audio, can be "over-specified"—containing redundant information. For example, chromagrams may include also traces of timbre and pitch range due to lower chroma bin resolution at lower frequencies. We measure the change in \emph{CLAP Score} across different conditioning setups to quantify how these added control conditions may override or "leak" into the backbone model’s original CLAP conditioning.

We compare three setups: 1) CLAP embeddings and control signals are \emph{aligned} (i.e., extracted from the same source audio signal); 2) CLAP embeddings and control signals are \emph{misaligned} (sourced from different audio examples); and 3) the model receives only the control signal, with CLAP conditioning omitted, labelled \emph{none}. Using these setups, we evaluate the following variables:

\noindent \textbf{Impact of Architecture:}
We compare all proposed LiLAC configurations as well as ControlNet to assess which architecture is more susceptible to CLAP leakage—where over-specified control signals (e.g., chroma) can dominate or obscure CLAP’s condition. By contrasting performance in \emph{aligned} vs. \emph{misaligned} configurations, we quantify how strongly each model prioritises information from the new control signals over pre-existing CLAP embeddings.

\vspace{2mm}
\noindent \textbf{Impact of Control Signal Specificity:} We evaluate three control signal types: (1) \emph{Chroma}, an over-specified signal containing pitch, timing, and timbral information; (2) \emph{Thresholded Chroma}, a variant where amplitudes $\geq 0.9$ are clipped to 1 and all others to 0, reducing fine timbral details to minimise leakage; and (3) \emph{Chords}, a well-specified signal conveying only harmonic structure. By analysing the difference in CLAP Score across these conditions, we quantify whether reducing surplus information in the conditional signal leads to more faithful adherence to the CLAP embedding.

\subsection{Subjective Evaluation}\label{sec:sa}

To subjectively evaluate LiLAC's effectiveness in conditioning the diffusion model, we conduct two listening tests loosely following MUSHRA guidelines \cite{MUSHRA} (i.e., using the true accompaniment as \emph{hidden reference} in our case). In both tests, participants rate sets of audio samples on a 100-point scale (0: poor, 100: excellent) based on two criteria: audio quality and subjective adherence, respectively, for each questionnaire.\footnote{\url{https://linktr.ee/lilactests}} Each test comprises 10 questions, with each presenting 5 unlabeled, randomly ordered samples corresponding to LiLAC$^\text{H}$, LiLAC$^\text{HTR}$, ControlNet, the ground-truth reference track and a negative anchor. Further details about each test are given below.

\subsubsection{Subjective Audio Quality (SAQ):} 

The SAQ test evaluates whether the control models audibly degrade the output quality. Participants are asked to rank each set of examples based on perceived audio quality and the presence of artifacts. For each question round, we start from a reference track, randomly sampled from the validation set, and extract the chromas and CLAP embeddings. Using these as conditioning, we generate an example with each of the three evaluated models. To create the negative anchor, we apply heavy compression (16 kbps MP3) to the ground truth track. The ground truth audio is used as a positive anchor. Shared chroma and CLAP conditioning in this test ensures that all examples are comparable, as they should ideally exhibit the same pitch distribution and timbre. This approach encourages participants to focus solely on evaluating audio quality.

\subsubsection{Subjective Condition Adherence (SCA):} 

The SCA test evaluates the models’ ability to adhere to chord conditioning, chosen for its flexibility in generating varied outputs (e.g., chords, bass lines, melodies). Participants rank how well the generated output aligns harmonically with the chord conditioning extracted from the multitrack recording. For each round, we begin with a multitrack recording and isolate one instrument track as the reference. Before generation, we evaluate the remaining multitrack; if it does not contain at least one polyphonic voice throughout the test sample, then the sample lacks sufficient harmonic content to reliably discern chords and is excluded from the listening test \. We extract chord symbols from the full multitrack and CLAP embeddings from the reference track, then generate an example for each tested model. The negative anchor is generated using the CLAP-conditioned Diff-A-Riff backbone, i.e., without chord or context conditioning. To facilitate comparison, the generated output is panned to the left, while the remaining multitrack is panned to the right.

\section{Results}
\label{sec:results}
In what follows, we present and discuss the results of the evaluations outlined in Sec.~\ref{sec:experiments}.

\begin{table}[h!]
    \small
    \centering
    \begin{tabular}{llccc}
        \toprule
        Model && APA $\uparrow$ & FAD $\downarrow$ & cMSE $\downarrow$ \\ 
        \toprule
        \multicolumn{2}{l}{Diff-a-Riff* \tiny{(432M)}} & 0.63 & 0.812 & 0.206 \\
        \multicolumn{2}{l}{\quad \small{+\,\emph{Context}}} & 1.00 & 0.508 & 0.148 \\ \midrule
        \multicolumn{2}{l}{\quad \small{+\,ControlNet} \tiny{(165M)}} & 1.00 & 0.507 & \textbf{0.052} \\ \midrule
        \multirow{4}{*}{\quad \small{+\,LiLAC}}
        & \small{H} \tiny{(32M)} & 1.00 & \textbf{0.506} & 0.057 \\
        & \small{HT} \tiny{(48M)} & 1.00 & 0.507 & 0.056 \\
        & \small{HR} \tiny{(47M)} & 1.00 & 0.509 & 0.056 \\ 
        & \small{HTR} \tiny{(64M)} & 1.00 & 0.507 & \underline{0.055} \\ \midrule
        \multicolumn{2}{l}{\quad \small{+\,LiLAC*} \tiny{(47M)}} & 1.00 & 0.508 & 0.070 \\ \bottomrule
    \end{tabular}
    \caption{Objective metrics: \emph{Audio Prompt Adherence} (APA) and \emph{Fréchet Audio Distance} (FAD) for audio quality; \emph{chroma Mean Squared Error} (cMSE) for chroma adherence. All models include CLAP conditioning except for Diff-A-Riff$^*$. Additionally, all new control models are conditioned on chroma.}
    \label{tab:ablation}
\end{table}

\subsection{Objective Experiments}

Table~\ref{tab:ablation} shows objective comparison between LiLAC and baseline models (see Sec.\ref{sec:cab}). The \emph{APA} metric—a distribution-based score bounded in range $[0, 1]$\cite{apa}—shows that all models conditioned on CLAP embeddings and chroma outperform the context-free Diff-A-Riff baseline (\emph{Diff-A-Riff$^*$}) and match the performance of the context-conditioned variant (\emph{+\,Context}). This indicates that chromagram-based conditioning is consistently effective at guiding alignment with the multitrack, regardless of architectural variations. Analogously, FAD scores remain comparable across all setups, suggesting that adding conditioning does not degrade audio quality. In particular, the near-identical FAD scores between \emph{Diff-A-Riff\,+\,Context} baseline, ControlNet and LiLAC models indicate that post-hoc conditioning doesn't compromise fidelity (low FAD, high APA) and may enhance controllability (high APA).

While these results provide a statistical perspective on the overall distribution of generated outputs and their alignment with musical context based on chroma alone, they do not directly reveal how faithfully models respond to specific chroma inputs. A pairwise metric is required to evaluate adherence on a per-example basis.
For this purpose, we report \emph{cMSE}. Here, ControlNet achieves the best absolute adherence to the chroma input, though all LiLAC configurations deliver competitive results while maintaining a considerably smaller parameter count (i.e., 165M in ControlNet versus 32M in LiLAC$^\text{H}$)\footnote{As the size of the adaptor blocks scales with the encoder's channel dimensions, the exact parameter reduction depends on the specific backbone architecture. For Diff-A-Riff \cite{diffariff}, our lightest configuration (LiLAC$^\text{H}
$) uses only 19\% of ControlNet's parameters.
This efficiency scales with larger architectures: for Stable Audio Open\cite{stableaudioopen}, the parameter usage drops to 10.2\%, while for image models such as SD V2\cite{SD}, it reduces dramatically to 2.6\%.}.
Among them, the HTR variant shows stronger performance, with \emph{cMSE} close to ControlNet. Interestingly, LiLAC*, a lightweight architecture inspired by prior work~\cite{sketch2sound}, exhibits limitations in controlling fine melodic nuances, thus, we exclude it from further evaluations.

\begin{table}[h]
    \small
    \centering
    \begin{tabular}{lccc}
        \toprule
        Model & Aligned & Misaligned & None\\
        \toprule
        Diff-a-Riff & 0.65 (0.17) & 0.65 (0.21) & 0.17 (0.21) \\ \midrule
        \quad \footnotesize{+\,LiLAC$^\text{H}$} & 0.67 (0.06) & \textbf{0.55} (0.06) & \textbf{0.57} (0.06) \\ 
        \quad \footnotesize{+\,LiLAC$^\text{HT}$} & 0.67 (0.06) & 0.54 (0.06) & 0.58 (0.05) \\
        \quad \footnotesize{+\,LiLAC$^\text{HR}$} & 0.67 (0.06) & 0.54 (0.06) & 0.58 (0.06) \\
        \quad \footnotesize{+\,LiLAC$^\text{HTR}$} & 0.67 (0.05) & 0.53 (0.06) & 0.58 (0.05) \\ \midrule
        \quad \footnotesize{+\,ControlNet} & 0.67 (0.05) & 0.52 (0.06) & 0.60 (0.05) \\ \bottomrule
    \end{tabular}
    \caption[Caption for LOF]{CS$\uparrow$\textsuperscript{a} (cMSE$\downarrow$) for models trained with chroma conditioning and Diff-A-Riff baseline. Each column corresponds with a different setting: \emph{aligned} pairs of chromagram and CLAP embedding, misaligned pairs, or without CLAP embedding (see Sec.~\ref{sec:cc}). 
    }
    \vspace{-0.32cm}\hspace{+5.65cm}\footnotesize\textsuperscript{a}CS$\downarrow$ in the \emph{None} case

    \label{tab:clap_models}
\end{table}
We next investigate how the new chroma adaptors interact with the backbone’s CLAP default conditioning (see Sec.\ref{sec:cc}). Table \ref{tab:clap_models} reports CLAP similarity scores (CS) and chroma reconstruction error (cMSE) across various model configurations. The results indicate a degree of information overlap between CLAP embeddings and chroma conditions. When both are present—but convey conflicting information—CLAP similarity scores noticeably decline, suggesting that chroma conditioning can partially overwrite or interfere with the original CLAP guidance. However, this drop is moderate, and the model still reconstructs a faithful chromagram, as reflected in the low cMSE even in the misaligned setting. This implies that the model makes reasonable sense of the conflicting conditionals—following the CLAP guidance as much as possible while still producing generations that faithfully adhere to the chroma conditioning.

Interestingly, in the CLAP-unconditioned setting, cMSE remains low while CS stays high. This supports our hypothesis of information redundancy between chromagrams and CLAP embeddings. It suggests that the model can still infer high-level cues—such as timbre—from chromagrams alone, allowing it to produce content that aligns with what CLAP, even in its absence. That said, the model's ability to maintain reasonable CLAP similarity without explicit CLAP input does not necessarily mean it reconstructs the original instrument associated with the chromagram. More likely, it infers a plausible instrument class—such as strings or pads—that fits the given pitch range and harmonic content. This suggests that the model interprets chroma conditioning sensibly: when CLAP is available, it may refine or overwrite the inferred instrument identity; when absent, it defaults to generating something timbrally coherent that aligns with the chroma structure.

Among all setups, the LiLAC architecture demonstrates the least susceptibility to CLAP interference, with its lightweight variant performing best under both conflicting and CLAP-ablated conditions. This robustness is likely due to its simplified structure, which limits the extent to which the model encodes auxiliary information alongside the CLAP embedding.

\begin{table}[h]
    \small
    \centering
    \begin{tabular}{lccc}
        \toprule
        Condition & Aligned & Misaligned & None \\ \toprule
        Diff-a-Riff & 0.65 (0.17) & 0.65 (0.21) & 0.17 (0.21) \\ \midrule
        \quad \small{+\,Chord} & 0.65 (0.14) & \textbf{0.64} (0.20)& \textbf{0.36} (0.18) \\
        \quad \small{+\,Thresh} & 0.65 (0.08) & 0.62 (0.17) & 0.45 (0.10) \\
        \quad \small{+\,Chroma} & \underline{0.67} (0.06) & 0.55 (0.06) & 0.57 (0.06) \\ \bottomrule
    \end{tabular}
    \caption{CS$\uparrow$\textsuperscript{a} (cMSE$\downarrow$) for LiLAC$^\text{H}$ trained on chord, thresholded chroma, and chroma with the Diff-a-Riff baseline (see Sec.~\ref{sec:cc}).}
    \vspace{-0.32cm}\hspace{+5.65cm}\footnotesize\textsuperscript{a}CS$\downarrow$ in the \emph{None} case
    \label{tab:clap_conds}
\end{table}

To reduce redundancy with CLAP, we evaluate LiLAC$^\text{H}$—our lightest and best-performing variant—using compressed melodic inputs: thresholded chromagrams (\emph{thresh}) and chords. Table~\ref{tab:clap_conds} compares performance across CLAP-aligned, misaligned, and unconditioned (\emph{None}) settings (see Sec.~\ref{sec:cc}).

Intuitively, less specific conditioning signals should exhibit lower overlap with CLAP embeddings. This is confirmed by the results: thresholding the chromagram significantly reduces the conflict between the two modalities. Among the tested conditions, chords show the least interference, with almost no drop in CLAP similarity in the misaligned case. Notably, even without CLAP conditioning, the chord model achieves a CLAP score of 0.36—up from 0.17 in the unconditioned baseline—suggesting that a substantial amount of global musical information (e.g., tempo, tonality) is implicitly captured by the chord input alone.

However, this robustness may come at the cost of reduced influence from the conditioning. We observe that weaker conditions, like chords, are more easily overridden by CLAP. For instance, the cMSE rises sharply from 0.08 to 0.17 when CLAP is misaligned, despite the sparsity of the chord-based input. This indicates that even minimal shifts in CLAP can significantly distort the resulting chroma, especially when the conditioning is less constraining. To confirm this, we re-evaluated the cMSE under the aligned setting for chords and observed a consistent value around 0.14. This validates that the observed jump in cMSE under misalignment reflects a genuine interference effect, rather than noise or sampling variance.

Overall, these findings highlight that our proposed conditioning mechanisms—especially when simplified—can effectively cooperate with CLAP guidance. When signals are aligned, they yield coherent generations that satisfy both timbral and harmonic constraints. When conflicting, the model tends to prioritise chroma while still leveraging CLAP to guide plausible generation. Even in the absence of CLAP, the model retains the ability to infer global cues such as tempo, tonality, and timbre from abstract melodic inputs, demonstrating the versatility and robustness of the conditioning approach.

\begin{table}[h]
    \small
    \centering
    \begin{tabular}{lcc}
        \toprule
        Model & SAQ $\uparrow$ & SCA $\uparrow$ \\ \toprule
        Reference & 61.5 $\pm$ 4.7  & 82.3 $\pm$ 4.3 \\ \midrule
        ControlNet & 56.4 $\pm$ 4.2 & 66.5 $\pm$ 5.1 \\
        LiLAC$^\text{H}$ & \textbf{60.6} $\pm$ 4.6 & 65.9 $\pm$ 4.8 \\
        LiLAC$^\text{HTR}$ & 58.7 $\pm$ 4.2 & \textbf{68.6} $\pm$ 4.9\\ \midrule
        Anchor & 26.4 $\pm$ 4.9 & 12.5 $\pm$ 2.5 \\ \bottomrule
    \end{tabular}
    \caption{Listener ratings with 95\% confidence intervals for SAQ and SCA across evaluated models (see Sec.~\ref{sec:sa}).}
    \label{tab:subject}
\end{table}

\subsection{Subjective Experiments}
Table~\ref{tab:subject} presents results from our Subjective Audio Quality (SAQ) and Subjective Condition Adherence (SCA) evaluations (see Sec.~\ref{sec:sa}). Across both tests, we collected a total of 1,250 ratings from 25 participants (11 for SCA and 14 for SAQ). In the SAQ test, all three models were rated comparably in terms of audio quality. The two LiLAC variants—especially the lightweight configuration—slightly outperformed ControlNet based on both mean score and average rank. However, these differences did not reach statistical significance, either among the three models or relative to the original stem, as indicated by the Friedman test ($p > 0.10$). This aligns with expectations: given that the backbone is frozen, it is unlikely that control mechanisms alone would improve audio quality. Nevertheless, the slightly better performance of LiLAC models may reflect a better preservation of pre-trained features, due to their smaller and more focused adapter layers.

The SCA results show similar outcomes. All models significantly outperform unconditional generation ($p < 0.0001$), and again show remarkable similarity to one another ($p > 0.8$). Regarding conditional adherence to the original stem, there is a more noticeable gap to the reference, however LiLAC$^{\text{HTR}}$ comes closer than other models, showing only marginally significant differences ($p = 0.09$).

Overall, these findings are consistent with our expectations: while audio quality remains stable across control methods, condition adherence benefits more clearly from architectural improvements like LiLAC$^{\text{HTR}}$.

\section{Conclusion}\label{sec:conclusion}

In this paper, we introduced a lightweight and flexible control methodology for text-to-audio diffusion models, inspired by ControlNet but with a significantly reduced parameter count. Our approach supports multiple configurations to accommodate control signals of varying complexity, while maintaining— and in some cases improving upon—the audio quality and condition adherence of the original method, as demonstrated in both objective and subjective evaluations. We believe this modular and efficient framework paves the way for more expressive and musically useful control modalities in audio generation.

As future work, we plan to explore the modularity and composability of multiple LiLAC controllers operating over a shared backbone, as well as investigate whether our method generalises to non-convolutional architectures such as DiT\footnote{Subsequent to our initial submission, we have successfully applied this architecture to the DiT-based Diff-a-Riff 2\cite{dar2} and seen similar results, suggesting that our methodology generalises well beyond convolutional architectures. For this we adapted the ControlNet methodology pioneered in \cite{pixartdelta} and again substituted the cloned transformer blocks for the pre-existing frozen blocks wrapped by our adaptor layers .}.

\section{Ethics Statement}
Sony Computer Science Laboratories is committed to exploring the positive applications of AI in music creation. We collaborate with artists to develop innovative technologies that enhance creativity. We uphold strong ethical standards and actively engage with the music community and industry to align our practices with societal values. Our team is mindful of the extensive work that songwriters and recording artists dedicate to their craft. Our technology must respect, protect, and honour this commitment.

LiLAC presents a lighter, more flexible control paradigm that enables musicians to exercise fine-grained control over generative model outputs, ensuring complete artistic agency. We hope the presentation of this model will encourage commercial text-to-audio model providers to incorporate similar additional controls into their workflows, thereby empowering users with greater creative autonomy.

Both LiLAC and its backbone model, Diff-A-Riff, have been trained exclusively on datasets that were legally acquired for internal research and development purposes. Consequently, neither the training data nor the models can be made publicly available. We remain committed to full legal compliance and proactively address all ethical considerations in our work.

\bibliography{ISMIRtemplate}

%
%
%
%

\end{document}